# Photoinduced Spin Centers in Photocatalytic Metal-Organic Framework UiO-66

[1*]Anastasiia Kultaeva, [2]Timur Biktagirov, [1]Andreas Sperlich, [1]Patrick Dörflinger, [3]Mauricio E. Calvo, [4]Eugenio Otal, [1]Vladimir Dyakonov

[1]*Experimental Physics 6 and Würzburg-Dresden Cluster of Excellence ct.qmat, Julius-Maximilian University of Würzburg, 97074 Würzburg, Germany*

[2]*University of Paderborn, Physics Department, D-33098 Paderborn, Germany*

[3]*Instituto de Ciencias de Materiales de Sevilla (Consejo Superior de Investigaciones Científicas-Universidad de Sevilla), C/Americo Vespucio, 49, Sevilla, 41092 Spain*

[4] *Research Initiative for Supra Materials, Shinshu University, 4-17- Wakasato, Nagano city, 380-8553, Japan*

*\*anastasiia.kultaeva@uni-wuerzburg.de*

**Abstract**
Metal-Organic frameworks (MOFs) are promising candidates for advanced photocatalytically active materials. These porous crystalline compounds have large active surface areas and structural tunability and are thus highly competitive with oxides, the well-established material class for photocatalysis. However, due to their complex organic and coordination chemistry composition, photophysical mechanisms involved in the photocatalytic processes in MOFs are still not well understood. Employing electron paramagnetic resonance (EPR) spectroscopy and time-resolved photoluminescence spectroscopy (trPL), the fundamental processes of electron and hole generation are investigated, as well as capture events that lead to the formation of various radical species in UiO-66, an archetypical MOF photocatalyst. A manifold of photoinduced electron spin centers is detected, which is subsequently analyzed and identified with the help of density-functional theory (DFT) calculations. Under UV illumination, the symmetry, g-tensors and lifetimes of three distinct contributions are revealed: a surface $O_2$-radical, a light-induced electron-hole pair, and a triplet exciton. Notably, the latter was found to emit (delayed) fluorescence. Our findings provide new insights into the photoinduced charge transfer processes, which are the basis of photocatalytic activity in UiO-66. This sets the stage for further studies on photogenerated spin centers in this and similar MOF materials.

## 1. Introduction

Photocatalysis, a process where light activates a catalyst to accelerate chemical reactions, plays a critical role in clean energy technologies, particularly in pollutant degradation and water splitting for hydrogen generation [1-4]. At its core, this process involves the generation of electron-hole pairs that drive reactions without consuming or altering the catalyst. In recent years, research has expanded to new classes of materials that offer enhanced control over the electron and hole generation, with Metal-Organic frameworks (MOFs) being at the forefront. MOFs are highly ordered, porous coordination polymers that combine metal ions or clusters with organic linkers, thereby exhibiting synergic properties among coordination and organic chemistry [5].

The relevance of MOFs in photocatalysis lies in their modular nature and unique tunability [6-8]. Their structural diversity allows designing materials with light absorption across a broad range of wavelengths, including the visible spectrum, which is crucial for maximizing the use of solar light in photocatalytic processes [9-11]. At the same time, their porous architecture with high surface-to-volume ratio, allows for efficient diffusion of reactant molecules to active sites, enhancing the overall photocatalytic performance. Finally, the design flexibility of MOFs enables the incorporation of active sites tailored for specific reactions, as well as the construction of composite materials that combine MOFs with other photocatalysts, such as semiconductors [4, 12].

A particularly noteworthy MOF in photocatalytic research and applications is UiO-66, which demonstrates UV responsiveness and remarkable stability under various conditions (Fig. 1A) [13, 14]. It is composed of $Zr_6(\mu_3\text{-O})_4(\mu_3\text{-OH})_4$ secondary building units (SBU) that consist of $\mu_3$-bridged $Zr^{4+}$ ions connected through terephthalate ligands. The introduction of chemical modifications, such as linkers containing functional groups, further enhances UiO-66's photocatalytic performance including the ability to shift its absorption spectrum to the visible range for more efficient light harvesting [9, 15, 16]. Like conventional solid-state photocatalysts, its activity relies on light-induced electron-hole separation, where these carriers are transferred to inner surface states or defective SBUs before reacting with adsorbed species (cf. Figure 1B) [15].

Despite the extensive studies of UiO-66, the fundamental mechanisms of charge separation and trapping within the material remain poorly understood. Electron paramagnetic resonance (EPR) spectroscopy, a well-established technique for studying charge traps in traditional solid-state photocatalysts such as $TiO_2$ [17, 18], has not yet yielded conclusive observations of stable photoinduced electron and hole states in pristine UiO-66, at least at temperatures above liquid nitrogen [19]. While localized states such as holes trapped at oxygen anions and electrons trapped at coordinatively unsaturated metal cations are expected to be accessible by EPR, delocalized electrons in the conduction band and holes in the valence band are EPR silent [20, 21]. Thus, this absence of conclusive EPR data suggests that the lifetimes of the trapped carriers in UiO-66 are short or that they have fast relaxation times that prevent their detection.

To address this knowledge gap, we employed broadband UV irradiation and low-temperature (T = 6 K) EPR spectroscopy to detect the elusive photoinduced electron and hole states in pristine UiO-66. Our results show that extended irradiation times are required to generate sufficient concentrations of spin centers detectable by EPR, even at low temperatures. Our experimental results indicate the presence of a manifold of photoinduced spin species, which we subsequently analyze and assign with the help of density functional theory (DFT) calculations. This work not only sheds light on the underlying photophysical processes in UiO-66 but also establishes a foundation for further investigations into the role of photogenerated spin centers in other MOF-based photocatalysts.

## 2. Results

As shown in Fig. 1C, broadband UV-irradiation of UiO-66 at low temperature (T = 6 K) results in the observation of a rich EPR spectrum that was previously unreported. The appearance of the spectrum suggests that it represents a superposition of various photogenerated paramagnetic centers. For the convenience of the following discussion, distinct groups of resonance signals constituting the spectrum are denoted as R0, R1, and R2.

The group R0 represents two or more partially overlapping lines situated around the free-electron $g$-factor ($g_e$ = 2.0023). Additionally, the R0 region contains the contribution of the broad EPR line also observed without irradiation ("dark" spectrum in Fig. 1C). The group R1 appears upon illumination and contains two partially resolved EPR lines, at $g$-factors of 2.044 and 2.036. These lines can either belong to two different paramagnetic species, or one spin center with distinct magnetic anisotropy. As R2, we denote two narrow, symmetric resonance lines with equivalent intensities and equidistantly separated from $g_e$. In addition, we observe a weak photoinduced EPR signal in the region of 155 mT ($g$ = 4.4), as shown in the inset to Fig. 1C, which can indicate the presence of a photoinduced triplet state (i.e., a center with the total spin $S$ = 1) in the sample [10].

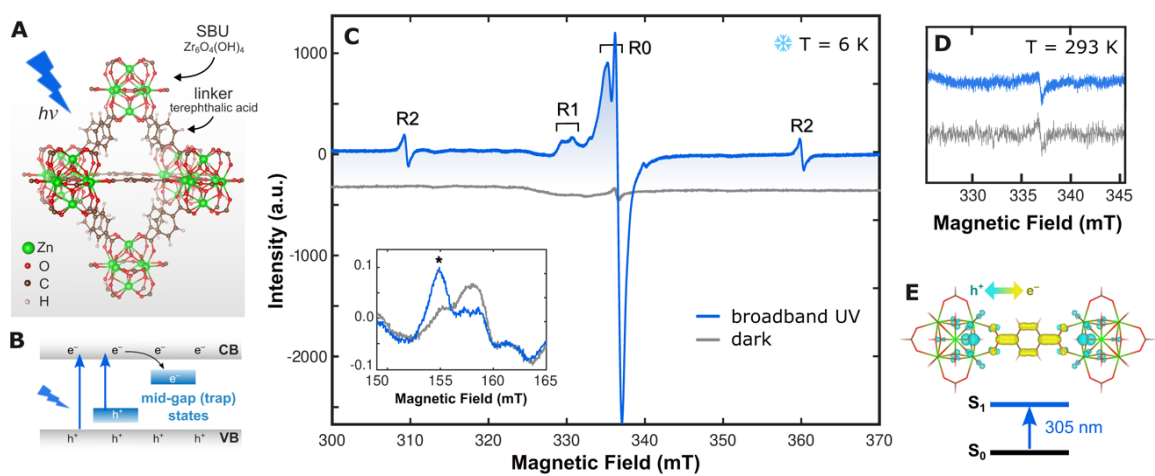

**Figure 1.** Detection of photoinduced spin centers in UiO-66. (A) Schematic atomic structure of UiO-66. (B) Scheme of UV-induced charge separation. Electrons are excited either from the valence band or from occupied mid-gap defect states near the valence band (VB) and can be subsequently trapped by defect or surface states near the conduction band (CB). The resulting localized electron and hole states can now be observed using EPR. (C) X-band CW EPR spectrum at T = 6 K upon broadband UV-irradiation (blue line) compared with the dark EPR spectrum measured at the same temperature (gray line). The inset shows the photoinduced EPR signal at ≈ 155 mT (asterisk). (D) UV illuminated (blue trace) and dark (gray trace) EPR spectra recorded at room temperature exhibit no presence of spin centers observed in (C). (E) TDDFT-calculated difference density (top) and excitation wavelength (bottom) of the lowest excited singlet state $S_1$ in UiO-66. The yellow isosurface shows the increase of electron density, while the cyan isosurface indicates its depletion. This equals a charge transfer in between linker (e−) and metal cluster (h+).

After the light is switched off, the photoinduced EPR signals remain stable at T = 6 K for at least several hours. However, after heating the sample back to room temperature, they disappear, allowing us to exclude their connection to photoionization damage of the material. Additionally, the X-ray diffraction data (Fig. S1) demonstrates that UV irradiation and low experimental temperatures do not cause the loss of crystallinity. Therefore, it is reasonable to assign the observed EPR signals to metastable radical species formed by light-induced separation and subsequent trapping (localization) of charge carriers. We use UV light with a broad spectrum, covering the range between 280 nm and 390 nm. According to our time-dependent DFT (TDDFT) calculations, this range covers the energy required to excite UiO-66 to the lowest singlet excited state ($S_1$; cf. Fig. 1E and Fig. S5 for UiO-66 absorption spectra).

The TDDFT-calculated transition density shown in Figures 1A and 1E indicates that photoexcitation to $S_1$ is accompanied by charge transfer from the oxygen atoms of the SBU to the linker. Therefore, it is a likely scenario that UV irradiation leads to the formation of electron-deficient oxygen-related radical species, which give rise to some of the resonance signals in the measured EPR spectrum. Specifically, the $g$-factors of the R1 signals (~2.04) remarkably agree with those reported for oxygen anions ($O^-$), commonly detected in metal oxide as a result of hole trapping by an oxygen atom at the surfaces. These species are characterized by axially symmetric $g$-tensors with the following relation between the axial ($g_{||}$) and in-plane ($g_\perp$) principal values: $g_\perp > g_{||} \approx g_e$. For instance, for MgO, a variety of $O^-$ centers with $g_\perp$ values in the range of 2.045 – 2.021 was observed [18]. A comparable range of $g$ values was predicted by our DFT calculation for the $O^-$ center (trapped hole) in the SBU of UiO-66 (cf. Figure S4, Supporting Information), supporting the tentative assignment of the R1 signals.

As for the R2 EPR lines, the splitting between them on the magnetic field axis of 50.6 mT (corresponding to 1.4 GHz) is too large to be attributed to $g$-factor anisotropy. It is also unlikely to be caused by electron-nuclear hyperfine splitting. While transition metal ions can exhibit large hyperfine splitting, the only magnetic isotope of Zr ($^{91}$Zr; natural abundance 11.2 %) has a nuclear spin $I$ = 5/2, which would result in six resonance transitions. A likely explanation of the splitting between R2 resonances is that it originates from the effect known as zero-field splitting (ZFS) typical for high-spin ($S$ > 1/2) states, such as triplets ($S$ = 1). ZFS can be as large as a few GHz even in organic systems. As we already mentioned, the EPR signal which is marked by an asterisk in Fig. 2D (see also the inset to Fig. 1C) can indicate a photoinduced triplet state, for which a half-field EPR signal corresponding to $\Delta m_S$ =±2, where $m_S$ is the spin projection quantum number, the so-called spin-forbidden transition is also expected. Thus, it can originate from the same triplet spin center as the R2 lines.

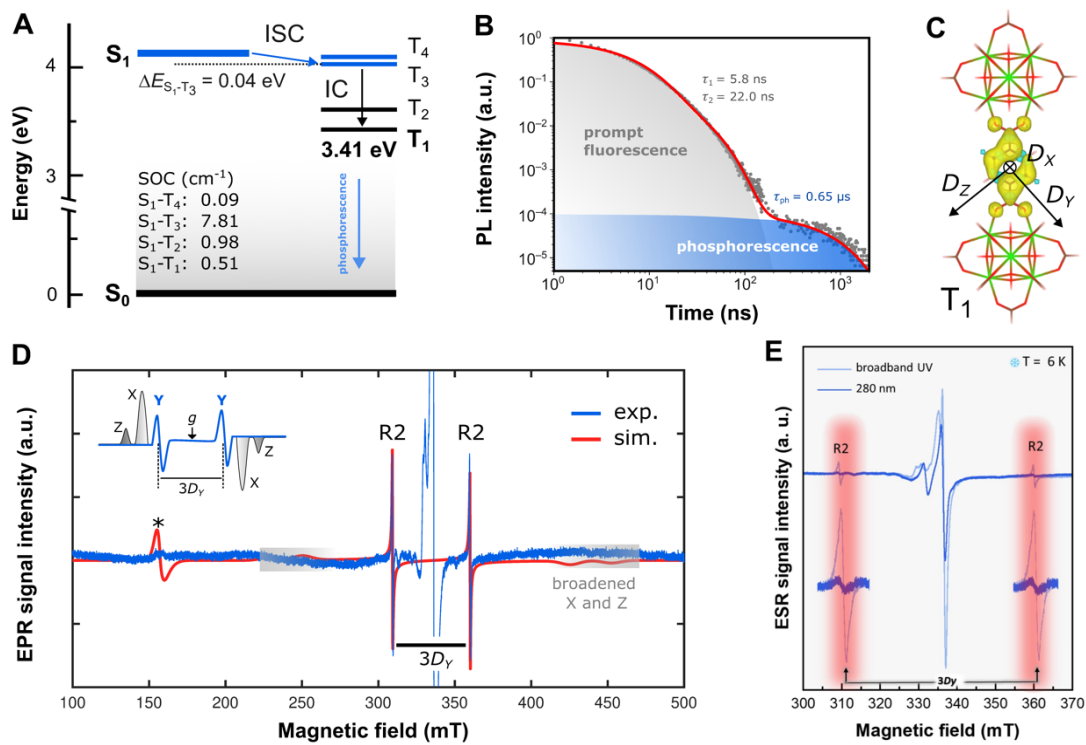

**Figure 2. Identification of a triplet state in defect-free UiO-66.** (A) TDDFT calculated energy level diagram and the corresponding SOC constants. Feasible excited states for singlet-to-triplet transformation are highlighted in blue. (B) Time-resolved photoluminescence (PL) decay at 10 K with a prompt fluorescence stemming from singlet excited states (grey area) followed by a phosphorescence signal on microsecond timescale, associated with triplet excited states (blue area). (C) DFT calculated spin-density distribution of the lowest excited triplet state, $T_1$, along with the principal directions of the ZFS tensor, $D_X$, $D_Y$, and $D_Z$. (D) Proposed simulation of the R2 EPR lines, and assuming the ZFS anisotropy of the $T_1$ triplet with $D_Y$ = 465 MHz and distortion-induced strain applied in the $D_X$ and $D_Z$ directions. The asterisk marks the assumed half-field transition. The inset shows a typical powder-average EPR spectrum of a triplet with an anisotropic ZFS. The transitions labeled X, Y, and Z correspond to the single-crystal contributions with the external magnetic field aligned with $D_X$, $D_Y$, and $D_Z$, respectively. (E) Comparison of the EPR spectra of UiO-66 irradiated with broadband UV and with 280 nm.

To theoretically investigate possible mechanisms of a triplet center formation in defect-free UiO-66, we analyzed the results of our TDDFT. Fig. 2A shows the energy level diagram of the lowest singlet ($S_1$) and triplet ($T_1$-$T_4$) excited states, as well as the spin orbit coupling (SOC) values between $S_1$ and the triplet states. Typically, an excited triplet can be accessed via an intersystem crossing (ISC) from the low-lying excited singlet followed by the internal conversion (IC) within the triplet manifold. The ISC process involves a spin-forbidden (singlet-triplet) horizontal transition driven by efficient SOC between the singlet and triplet wavefunctions. As a rule of thumb, facile ISC channels require small singlet-triplet gap ($\Delta E_{S-T}$ < ±0.37 eV) and high SOC values (>0.04 meV or 0.3 cm$^{-1}$) [22, 23]. As shown in Fig. 2A, there are excited states in the triplet manifold (such as $T_3$) close to $S_1$ in energy and exhibiting strong SOC to facilitate efficient ISC.

Once populated via ISC, the excited triplet state can return to the ground state through a spin-forbidden radiative transition known as phosphorescence, particularly when the rates of non-radiative energy transfer into lattice vibrations are suppressed. Therefore, evidence for the presence of long-lived triplet excited states can be found in photoluminescence kinetics measured after pulsed excitation of singlet states (Fig. 2B and Fig. S5, Supporting Information). The photoluminescence decay curve of UiO-66 exhibits a prompt fluorescence from the singlet excited states (best fitted by a biexponential decay with time constants in the order of nanoseconds; cf. Fig 2B), followed by a delayed fluorescence signal on microsecond time scales – a typical indication of phosphorescence from the lowest triplet states.

The spin density distribution of the lowest triplet ($T_1$) is illustrated in Fig. 2C and is almost entirely localized on the organic linker of UiO-66. The principal values and principal directions of its ZFS tensor are determined predominantly by spin-spin interaction between unpaired electrons and, therefore, reflect the spatial distribution of the spin density [24]. Expectedly, the calculated ZFS tensor is highly anisotropic with the principal values $D_X$ = 1282 MHz, $D_Y$ = 785 MHz, and $D_Z$ = –2068 MHz (and the respective principal directions shown in Fig. 2C). The calculated $g$-tensor of the triplet is almost isotropic and close to free electron g-factor $g_e$.

As schematically illustrated in the inset of Fig. 2D, the ZFS anisotropy of the $T_1$ triplet state should manifest in six EPR lines. They originate from ensemble-averaging of the different molecular orientations with respect to an external magnetic field, denoted as X, Y, and Z, each of them exhibiting two $\Delta m_S$ = 1 resonance transitions. We do not observe the X and Z components – at least, at the field positions expected from the DFT calculated ZFS tensor. However, we can explore a scenario, in which the X and Z components are suppressed by orientation-dependent line broadening. Such broadening, often observed in CW EPR spectra of various organic and inorganic compounds, is conventionally described by the statistical distribution of the principal values of the $g$-tensor around their mean values and is referred to as "$g$-strain" [25]. For example, in organic nitroxide radicals, $g$-strain is most pronounced for the $g_X$ values and is explained by site-to-site local structural variations [26, 27].

The atomic structure of UiO-66 can exhibit a certain degree of variability (due to intrinsic dynamic flexibility causing local structural deformations [28]) and, therefore, can favor the observation of the $g$-strain. To illustrate this idea, Fig. 2D shows the simulated EPR spectrum of a triplet state with the $D_Y$ value fitted to the resonance transition fields of the R2 lines ($D_Y$ = 465 MHz), the $D_Z$ value is kept as in the DFT results, and 4% statistical variation (strain) is applied to the $g_X$ and $g_Z$ parameters. This relatively small $g$-strain can lead to the disappearance of the X and Z components from the ensemble-averaged (powder) EPR spectrum. It is important to note that we cannot exclude the possibility that the R2 signals are related to other paramagnetic species such as impurities in the pores of the MOF samples. However, our analysis strongly supports their assignment to the excited triplet state inherent to the structure of UiO-66, providing compelling evidence for this interpretation.

Finally, Fig. 2E compares the EPR spectra obtained by excitation with a broadband UV source and a target wavelength of 280 nm from a LED. While broadband excitation is expected to address a vast ensemble of mid-gap defect states and higher excited states, the 280 nm wavelength should more specifically excite the lowest singlet state ($S_1$ in Fig. 2A), facilitating the pathway of the triplet formation predicted by TDDFT. Thereby, selective photoexcitation allows the separation of the described triplet state from the other photoexcitable centers in the structure of UiO-66.

The EPR spectrum taken at 280 nm excitation exhibits substantial differences in its central region. In contrast, the R2 signal retains, although with reduced intensity, supporting its association with the triplet state. It is important to note that the power of the light sources is comparable and is therefore not responsible for the lower signal intensity of R2. Consequently, these findings strongly imply a significant role of mid-gap defect states in the photoexcitation process and the dynamics of photoexcited carriers within UiO-66.

Following the tentative assignment of the observed photogenerated spin centers, we aim to explore their thermal stability. For this purpose, we measured the temperature dependence of the EPR spectrum after the UV source was turned off (Fig. 3A). As mentioned above, the observed centers are stable at 6 K after irradiation. But due to their distinct chemical reactivity and electron spin relaxation characteristics, the R0, R1 and R2 centers are expected to exhibit varied behaviors upon heating. At first, the R2 resonance signal disappears between 15 K and 25 K. The R1 EPR signals become undetectable at ≈ 50 K. The activation energy or rather trapping energy of the spin centers R1 and R2 can therefore be very roughly estimated from the temperature behavior of their disappearance and should be on the order of magnitude of thermal energy $k_B T \approx 3$ meV. If we assume that R1 is associated with electron-deficient oxygen species (such as $O^-$ radicals, as discussed previously and supported by our DFT calculations), this behavior seems expected. In many metal oxides, the surface-bound $O^-$ radical is only observed at low temperatures [17].

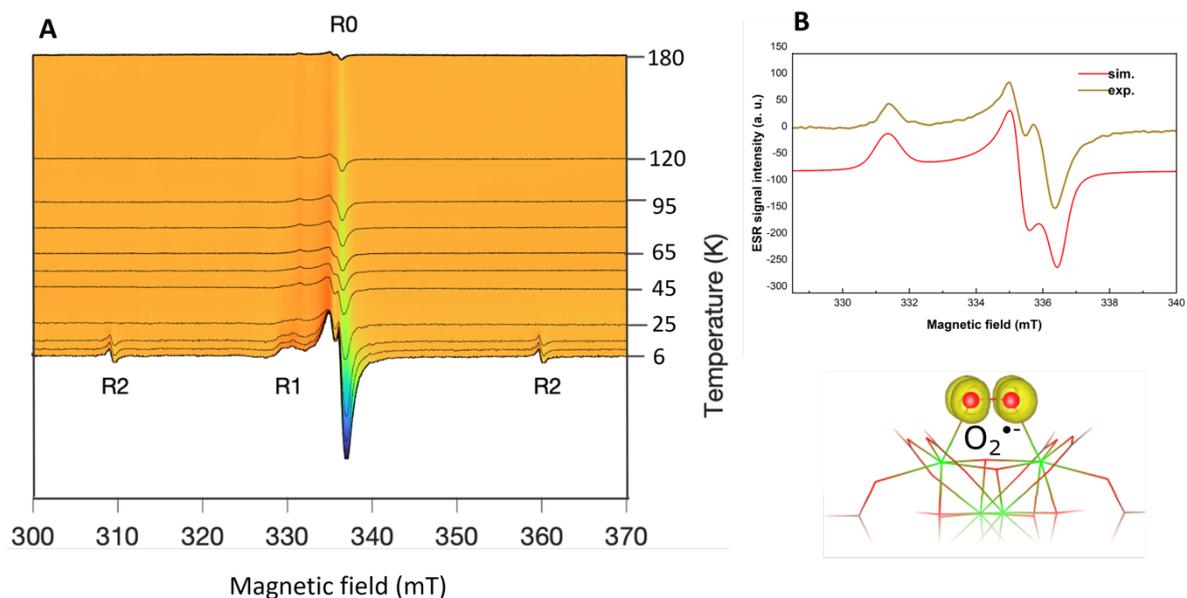

**Figure 3. Thermal stability of photoinduced radical species.** (A) Evolution of the EPR spectrum upon heating after the UV illumination is switched off. (B) The EPR spectrum observed at 180 K in (A), which is attributed to the superoxide radical adsorbed at the unsaturated metal site (see the structural model shown below) and simulated using the anisotropic $g$-factor discussed in the main text.

The temperature dependence of R1 and R2 explains why these EPR signals were never previously observed. Our results show that sufficiently low temperatures (T < 15 K) are required to reach a detectable signal of the photoinduced radical species in pristine UiO-66. However, most of the earlier studies were performed at room temperature, with a few going down to liquid nitrogen temperatures (T > 77 K). This is understandable since some chemical modifications of UiO-66 were reported to exhibit photoinduced EPR spectra already at room temperature. For example, $NH_2$-UiO-66, a modified version of UiO-66 incorporating 2-aminoterephthalic acid as the organic linker, exhibits a photoinduced EPR signal at room temperature that has been interpreted as a superoxide radical, $O_2^{\bullet-}$ [29]. It forms when photogenerated electrons reduce molecular oxygen adsorbed at the unsaturated $Zr^{4+}$ sites of a defective SBU.

Superoxide radicals are key intermediates in photocatalytic reactions, such as the degradation of organic pollutants, making the detection of this $O_2^{\bullet-}$ signal a clear indicator of efficient electron transfer to oxygen and the resulting catalytic activity in $NH_2$-UiO-66. However, this signal has not been previously observed in pristine UiO-66. Aside from the presence of an –$NH_2$ group at the linker, which facilitates better electron-hole separation and lowers the band gap in $NH_2$-UiO-66, the two MOFs share the same architecture and same SBU. This raises the question of why charge transfer to oxygen is not detected in UiO-66, even if this process occurs at lower efficiency.

Here, by analyzing the temperature dependence shown in Figure 3A, we found that an EPR signal resembling that in $NH_2$-UiO-66 appears in UiO-66 upon the thermally induced decay of the other species (above 45 K). This signal, recorded at 180 K, is presented in Figure 3B along with its simulation. It can be well described using an anisotropic *g*-factor with the principal values $g_X$ = 2.002, $g_Y$ = 2.009, and $g_Z$ = 2.033, which align well with our DFT calculations for an $O_2^{\bullet-}$ radical adsorbed at the missing-linker site of SBU: $g_X$ = 2.0037, $g_Y$ = 2.0109, $g_Z$ = 2.0264. Most importantly, when we measure the photogenerated EPR signal in $NH_2$-UiO-66 at room temperature and overlay it with the 180 K spectrum of UiO-66, we observe a near-exact match in the resonance transitions, signifying the same principal *g*-values (see Figure S6, Supporting Information). This strongly suggests that the same type of $O_2^{\bullet-}$ spin center is formed in both MOFs, despite the different conditions under which they are observed.

This observation provides a missing piece in the understanding of the comparative photocatalytic performance of UiO-66 and its modified versions. It provides solid evidence that fundamental charge transfer mechanisms underlying the photocatalytic activity of these MOFs are not inherently different**.** Both UiO-66 and $NH_2$-UiO-66 can generate radical species, such as superoxide, but functionalization with electron donating groups like –$NH_2$ is effective in enhancing performance at ambient conditions. From a practical perspective, this insight can guide the rational design of future MOFs for photocatalytic applications. First, it provides visual evidence that the –$NH_2$ group in $NH_2$-UiO-66 does not only decrease the band gap, but also reduces the activation energy needed for electron transfer and stabilizes the resulting charges, enabling these reactions to occur efficiently at higher temperatures. Second, the results here suggest that even unmodified UiO-66 could be activated under specific conditions, such as low-temperature, to potentially achieve similar catalytic outcomes. This, in turn, paves the way for utilizing UiO-66 in applications where these conditions are feasible or desirable.

## 3. Conclusions

Our work advances the understanding of the photophysics of MOF materials by revealing previously undiscovered details on photoinduced charge transfer, separation, and capture within defect-free UiO-66, an archetypical MOF photocatalyst. We demonstrate for the first time that photo-induced charge carriers can be detected in this MOF under specific conditions, namely cryogenic temperatures ($T < 20$ K) and sufficiently long UV-irradiation time. These conditions compensate for the thermal instability of the photogenerated spin species associated with their rapid recombination and/or fast electron spin relaxation. The detectability of trapped photoinduced charge carriers by means of EPR spectroscopy offers a promising tool for understanding and tuning the photocatalytic properties of this and other types of MOF materials. Therefore, by establishing the experimental conditions that favor the observation of these photogenerated spin centers, we provide the guidelines for subsequent investigations of MOF photocatalysts.

**Conflicts of interest**

There are no conflicts to declare.

**Supplementary Materials**

Supporting Information contains supplementary explanation and data, including XRD, PL, and UV-VIS analysis, additional EPR spectra and DFT simulations, light source analysis. SI includes Figures S1 – S8.

**Acknowledgements**

A.K. and V.D. acknowledge financial support from the Würzburg-Dresden Cluster of Excellence on Complexity and Topology in Quantum Matter ct.qmat (EXC 2147, DFG project ID 390858490). P.D. acknowledges the German Research Foundation (DFG) program SPP2196 under DY18/14-2 (Project number 424101351). The authors thank Paul Konrad for participating in the LED set-up preparation and Philipp Rieder for assisting with the XRD measurements.

**Data Availability**

The data that support the findings of this study are available from the corresponding author upon reasonable request.

**Authors contributions**

A.K. performed the magnetic resonance measurements and evaluated the data with help of A.S. and V. D., T. B. performed the DFT calculations. E. O. and M. C. performed the chemical synthesis and UV-VIS characterization. P.D. did PL measurements, XRD and characterization of light sources. All the authors contributed to analysis of the data, discussions and to the writing of the paper.

**ORCID**

Anastasia Kultaeva: 0000-0003-2040-7193
Timur Biktagirov: 0000-0002-6819-7124
Andreas Sperlich: 0000-0002-0850-6757
Patrick Dörflinger: 0000-0003-0872-3513
Mauricio E. Calvo: 0000-0002-1721-7260

Eugenio Otal: 0000-0003-0801-854X
Vladimir Dyakonov: 0000-0001-8725-9573

**References**

1. L. Candish, K. D. Collins, G. Cook, J. J. Douglas, A. Gómez-Suárez, A. Jolit, S. Keess, *Chem. Rev.* 2022, **122**(2), 2907.

2. P. Bellotti, H.-M. Huang, T. Faber, F. Glorius, *Chem. Rev.* 2023, **123**(8), 4237.

3. J. Twilton, C. C. Le, P. Zhang, M. H. Shaw, R. W. Evans, D. W. C MacMillan, *Nat. Rev. Chem.* 2017, 1(1), 52.

4. S. Navalón, A. Dhakshinamoorthy, M. Álvaro, B. Ferrer, H. García, *Chem. Rev.* 2022, **123**(1), 445.

5. Q. Wang, D. Astruc, *Chem. Rev.* 2020, **120**(2), 1438.

6. H.-C. Zhou, J. R. Long, O. M. Yaghi, *Chem. Rev.* 2012, **112**(2), 673.

7. M. O'Keeffe, O. M. Yaghi, *Chem. Rev.* 2012, **112**(2), 675.

8. N. Stock, S. Biswas, *Chem. Rev.* 2012, **112**(2), 933.

9. E. H. Otal, M. L. Kim, M. E. Calvo, L. Karvonen, I. O. Fabregas, C. A. Sierra, J. P. Hinestroza, *Chem. Comm.* 2016, **52**, 6665.

10. Y. Fu, D. Sun, Y. Chen, R. Huang, Z. Ding, X. Fu, Z. Li, *Ang. Chem.* 2012, **51**(14), 3364.

11. C. H. Hendon, D. Tiana, M. Fontecave, C. Sanchez, L. D'arras, C. Sassoye, L. Rozes, C. Mellot-Draznieks, A. Walsh, *J. Am. Chem. Soc.* 2013, **135**(30), 10942.

12. Q. Wang, Q. Gao, A. M. Al-Enizi, A. Nafady, S. Ma., *Inorg. Cem. Front.* 2020, **7**, 300.

13. J. Winarta, B. Shan, S. M. Mcintyre, L. Ye, C. Wang, J. Liu, B. Mu, *Cryst. Growth Des.* 2020, **20**(2), 1347.

14. J. H. Cavka, S. Jakobsen, U. Olsbye, N. Guillou, C. Lamberti, S. Bordiga, K. P. Lillerud, *J. Am. Chem. Soc.* 2008, **130**, 13850.

15. A. De Vos, K. Hendrickx, P. Van Der Voort, V. Van Speybroeck, K. Lejaeghere, *Chem. Mater.* 2017, **29**(7), 3006.

16. Y. He, C. Li, X.-B. Chen, Z. Shi, S. Feng, *ACS Appl. Mater. Interfaces* 2022, **14**(25), 28977.

17. T. Berger, M. Sterrer, O. Diwald, E. Knözinger, D. Panayotov, T. L. Thompson, J. T. Yates, *J. Phys. Chem. B* 2005, **109**(13), 6061.

18. M. Chiesa, M. C. Paganini, S. Livraghi, E. Giamello, *PCCP* 2013, **15**(24), 9435.

19. M. Zeama, M. Morsy, S. Abdel-Azeim, M. Abdelnaby, A. Alloush, Z. Yamani, *Inorganica Chim. Acta* 2020, **501**, 119287.

20. E. Carter, A. F. Carley, D. M. Murphy, *J. Phys. Chem. C* 2007, **111**(28), 10630.

21. G., Pinarello, C. Pisani, A. D'Ercole, M. Chiesa, M. C. Paganini, E. Giamello, O. Diwald, *Surf. Sci.* 2001, **494**(2), 95.

22. M. J. Leitl, V. A. Krylova, P. I. Djurovich, M. E. Thompson, H. Yersin, *J. Am. Chem. Soc.* 2014, **136**(45), 16032.

23. Y. Tao, R. Chen, H. Li, J. Yuan, Y. Wan, H. Jiang, C. Chen, Y. Si, C. Zheng, B. Yang, G. Xing, W. Huang, *Adv. Mater.* 2018, **30**(44), 1803856.

24. S. Richert, A. E. Tait, C. R. Timmel, *J. Magn. Reson.* 2017, **280**, 103.

25. W. R. Hagen, *J. Magn. Reson. (1969)* 1981, **44**(3), 447.

26. A. Nalepa, K. Möbius, M. Plato, W. Lubitz, A. Savitsky, *Appl. Magn. Reson.* 2018, **50**, 1.

27. A. Savitsky, A. Nalepa, T. Petrenko, M. Plato, K. Möbius, W. Lubitz, *Appl. Magn. Reson.* 2022, **53**, 1239.

28. J. Hajek, C. Caratelli, R. Demuynck, K. De Wispelaere, L. Vanduyfhuys, M. Waroquier, V. Van Speybroeck, *Chem. Sci.* 2018, **9**(10), 2723.

29. J. Long, S. Wang, Z. Ding, S. Wang, Y. Zhou, L. Huang, X. Wang, *Chem. Comm.* 2012, **48**(95), 11656.

Supporting Information for

# Photoinduced Spin Centers in Photocatalytic Metal-Organic Framework UiO-66

[1*]Anastasiia Kultaeva, [2]Timur Biktagirov, [1]Andreas Sperlich, [1]Patrick Dörflinger, [3]Mauricio E. Calvo, [4]Eugenio Otal, [1]Vladimir Dyakonov

[1]*Experimental Physics 6 and Würzburg-Dresden Cluster of Excellence ct.qmat, Julius-Maximilian University of Würzburg, 97074 Würzburg, Germany*

[2]*University of Paderborn, Physics Department, D-33098 Paderborn, Germany*

[3]*Instituto de Ciencias de Materiales de Sevilla (Consejo Superior de Investigaciones Científicas-Universidad de Sevilla), C/Americo Vespucio, 49, Sevilla, 41092 Spain*

[4] *Research Initiative for Supra Materials, Shinshu University, 4-17- Wakasato, Nagano city, 380-8553, Japan*

**METHODS**

**Sample preparation**
UiO-66 was synthesized using a modified procedure previously reported by Katz et al. [1S]. In a 500 ml Schott Duran flask, 300 ml of DMF and 20 ml of concentrated HCl were mixed under a fume hood. To this solution, 2.51 g of $ZrCl_4$ was added in small portions, also under the fume hood. The solid dissolved quickly, and then 2.49 g of terephthalic acid was added to the resultant solution. To facilitate the dissolution of terephthalic acid, the mixture was subjected to an ultrasonic bath for 30 minutes. After complete dissolution, the screw-cap bottles were sealed and maintained at 120°C for 12 hours. The solids were isolated by centrifugation and washed twice with DMF and $CH_2Cl_2$.

**Continuous wave electron paramagnetic resonance**
CW EPR was done on a commercial Magnettech spectrometer MS5000 outfitted with an Oxford ESR 900 He flow cryostat. The deviation of the temperature was less than 0.2 K. Two types of light sources were used, a broadband Dymax BlueWave 50 UV lamp (280 nm to 450 nm, see Fig. S7) and a LED source with a wavelength of 280 nm from Thorlabs. The characteristic wavelengths of high intensity of the broadband lamp are 310 nm, 370 nm, 400 nm and 440 nm. The power of the LED and UV lamp were measured considering the distance between the sources and the irradiated sample. Thus, the LED with a wavelength of 280 nm has a power of 6.5 mW at a distance of 11 cm at the sample, and the broadband UV lamp 5.1 mW respectively. A microwave power of 1 mW was chosen for optimal signal-to-noise ratio of the main EPR spectra without saturation effects. However, experiments on saturation behavior have also been conducted.

**EPR spectral simulations**

For EPR spectral simulations, we used the EasySpin software package based on MatLab [2S]. For the centers with the total spin $S$ = 1/2, the spin Hamiltonian included only the Zeeman interaction term expressed as:

$$H_Z = \beta_e \mathbf{SgB} = \beta_e[g_x S_x B_x + g_y S_y B_y + g_z S_z B_z] \qquad (1)$$

where $\beta_e$ is the Bohr magneton; $\mathbf{B}$ is the external magnetic field vector, $\mathbf{S}$ is the electron spin operator, and $\mathbf{g}$ is the is the electron g-tensor with the principal values $g_x$, $g_y$, and $g_z$. In case of a g-tensor, $g_x = g_y = g_\perp$ and $g_y = g_\parallel$.

In case of triplet ($S$ = 1) states, anisotropic magnetic dipole-dipole interaction between two unpaired electrons results in lifting of the degeneracy among the three spin sublevels, even without an external magnetic field. This phenomenon is known as the zero-field splitting (ZFS) interaction and can be described by the spin Hamiltonian:

$$H_{zfs} = \mathbf{SDS} = D_x S_x^2 + D_y S_y^2 + D_z S_z^2 = D\left(S_z^2 - \frac{1}{3}S^2\right) + E(S_x^2 - S_y^2) \qquad (2)$$

where $D_x$, $D_y$, $D_z$ are the principal values of the ZFS tensor $\mathbf{D}$. The ZFS interaction is typically specified by the two parameters $D = \frac{3}{2}D_z$ and $E = \frac{1}{2}(D_x - D_y)$ to best describe its traceless nature [3S, 4S].

**Steady-State Photoluminescence (SSPL) and Time-Correlated Single-Photon Counting (TCSPC)**

All photoluminescence (PL) measurements were performed using an FLS 980 spectrometer (Edinburgh Instruments). The samples were placed in a sample holder and measured in reflection geometry. The emission from the sample was collimated and detected by a photomultiplier tube positioned behind a monochromator. For steady-state measurements, the samples were excited with a 375 nm laser having a spot size of 100 µm in diameter, a fluence of approximately 88 nJ/cm² with a repetition rate of 50 ns. Time-Correlated Single Photon Counting (TCSPC) measurements were additionally also performed using a 375 nm laser. For the room temperature decay, a fluence of approximately 76 nJ/cm² with a repetition rate of 2 µs was used. The temperature-dependent decay measurements consisted of two merged decays, each with a fluence of approximately 31 nJ/cm² and a repetition rate of 100 µs. These decays differed only in the measured time window to optimize data distribution and acquisition time. To determine the charge carrier lifetime, the transient decay was fitted with a tri-exponential decay model for the temperature-dependent decay:

$$y(x) = -A_1 \cdot exp^{-x/t_1} + A_2 \cdot exp^{-x/t_2} + A_3 \cdot exp^{-x/t_3} \qquad (3)$$

where $A_1$, $A_2$, $A_3$ are the weighting factors that were used to analyze the contribution of exponential curves to the total decay and are equal to $2.46 \cdot 10^{-2}$, $6.21 \cdot 10^{-1}$ and $9.2 \cdot 10^{-5}$, respectively. The time constants are described by $t_1$, $t_2$ and $t_3$ and equal to 22 ns, 5.6 ns and 650 ns, respectively.

**DFT calculations**

The DFT calculations were performed with the ORCA program package [5S] using the TPSSh exchange-correlation functional [6S] and the def2-TZVP basis set [7S] combined with the RIJCOSX approximation and the corresponding auxiliary basis set [8S]. Both ground-state and excited state calculations were performed using a cluster model cut from the crystal structure of UiO-66. The cluster model consisted of one linker and two SBUs with the remaining (unsaturated) coordination sites of the SBUs terminated by $HCOO^-$ anions (see Table S1). During geometry optimization, the coordinates of the carbon atoms of each $HCOO^-$ were fixed. For the excited states, TDDFT calculations [9S] were used with the optimized ground-state atomic structures. The adiabatic TDDFT approach is known to have limitations when applied to transition metal complexes. However, since the unoccupied Zr d-states are located high in the conduction band (see the discussion in the main text), the lowest excited states of UiO-66 considered in this work are almost entirely associated with the organic moiety. Therefore, TDDFT is expected to provide reliable results in this context. For the calculation of the ZFS tensor, the lowest excited triplet state was modeled by constraining the spin multiplicity. The calculated ZFS included both the spin-spin and spin-orbit contributions. To diminish the spin contamination error in the spin-spin ZFS [10S], it was calculated using the unrestricted natural orbital (UNO) approach [11S] and the generalized gradient (GGA) based PBE functional [12S].

**X-ray diffraction**

X-Ray diffraction (XRD) measurements were performed using a General Electric XRD 3003 TT with a monochromatic Cu-Kα radiation source (U = 40 kV, I = 40 mA) with a wavelength λ of 1.5406 Å. The measurements were performed at room temperature. Simulations were performed with Diamond software. The structure has a cubic space group *Fm-3m* (225).

**ADDITIONAL RESULTS**

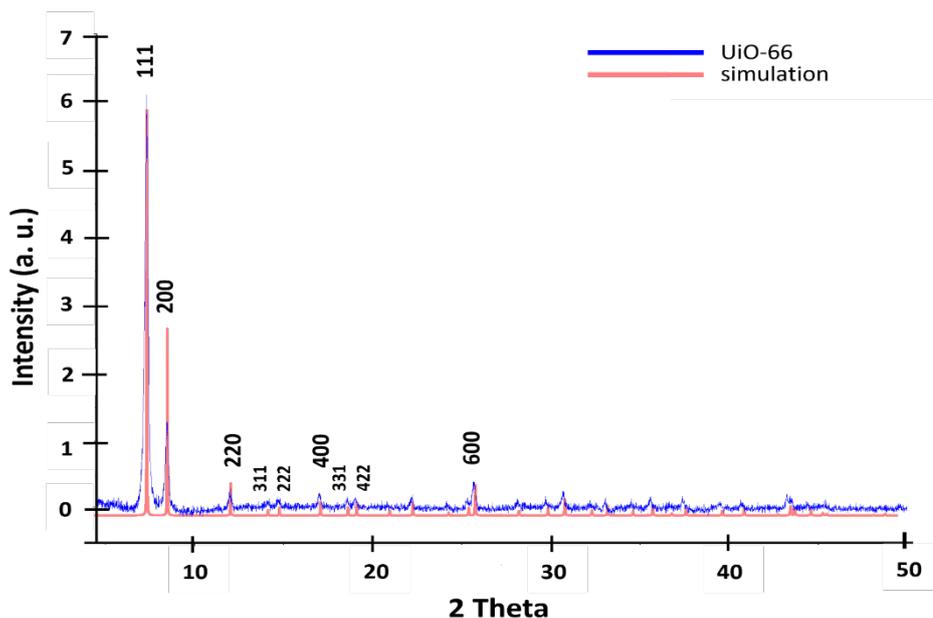

**Figure S1.** X-Ray diffraction pattern for UiO-66 powder sample after UV irradiation at T = 6 K and heating back to room temperature. No significant broadening of the peaks was found, indicating that the crystallinity of the structure was not lost after UV irradiation and low experimental temperatures.

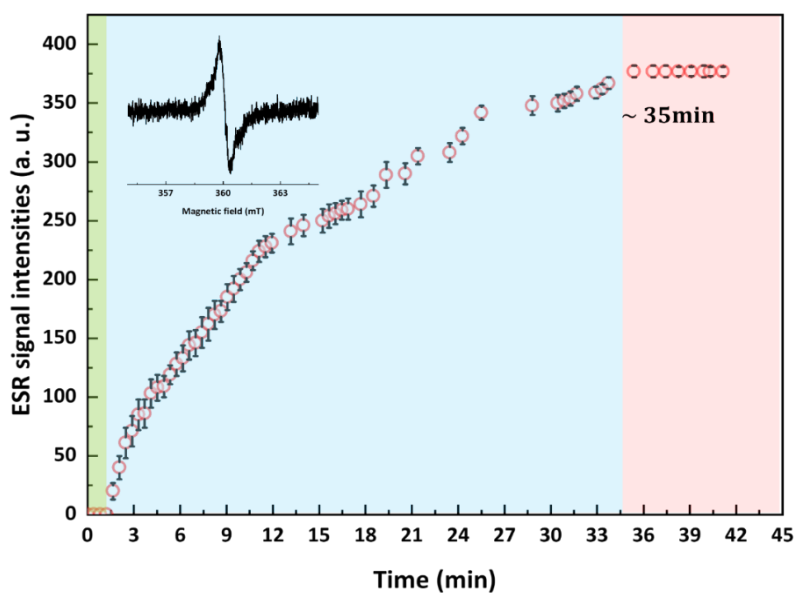

**Figure S2**. Time dependence of R2 signal intensity growth. One of the signal peaks corresponding to one of the $\Delta m_s = \pm 1$ transitions was chosen to measure the time dependence. The measurements were performed at T = 6 K with fixed modulation amplitude and microwave power for each scan. Also, the receiver gain remained the same throughout the experiment.

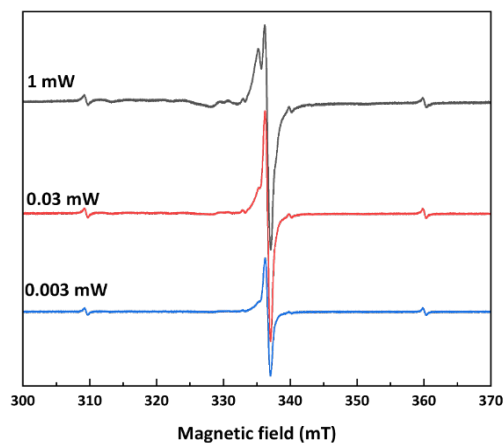

**Figure S3**. The microwave dependence of R0, R1 and R2 signals. The distinct behavior under microwave power saturation displays differences in spin-spin and spin-lattice relaxation times due to the different nature and coordination environment of the R0, R1, and R2 centers.

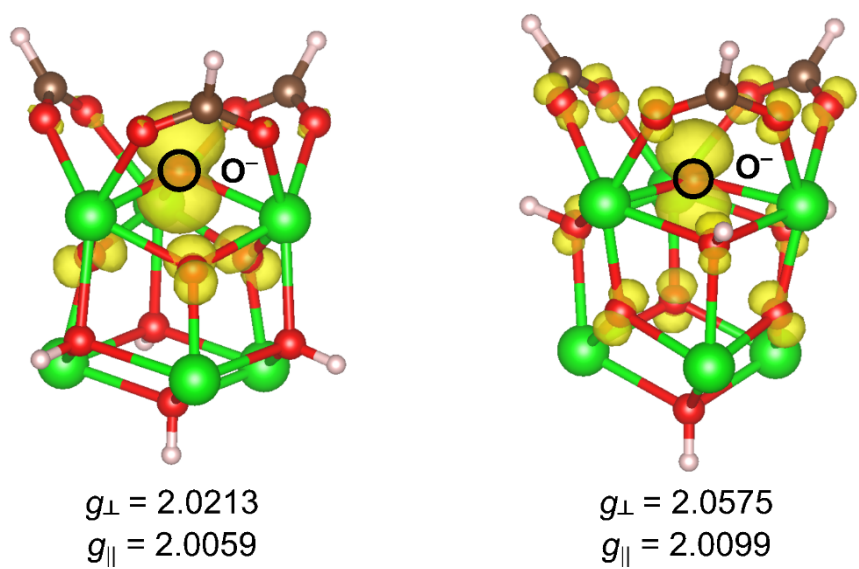

$g_\perp = 2.0213$
$g_\parallel = 2.0059$

$g_\perp = 2.0575$
$g_\parallel = 2.0099$

**Figure S4**. Two representative DFT models of a hole trapped at a $\mu_3$-O atom of the UiO-66 SBU (i.e., the O⁻ radical anion) with different arrangements of the surrounding $\mu_3$-O and $\mu_3$-OH groups. The electron spin densities are shown as yellow isosurfaces and the calculated principal values of the g-tensors are shown below each structure, illustrating the range of the principal *g*-values expected for this center.

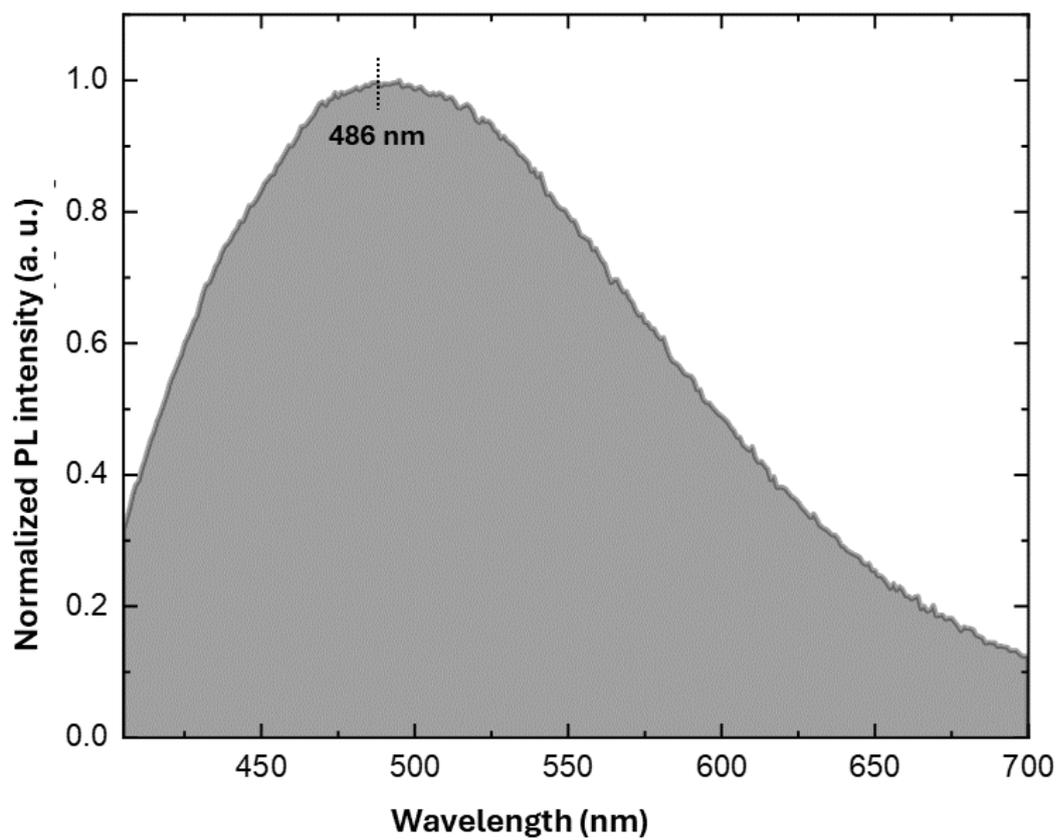

**Figure S5**. Photoluminescence spectrum of UiO-66 measured at 10 K. The maximum is located at approximately 486 nm.

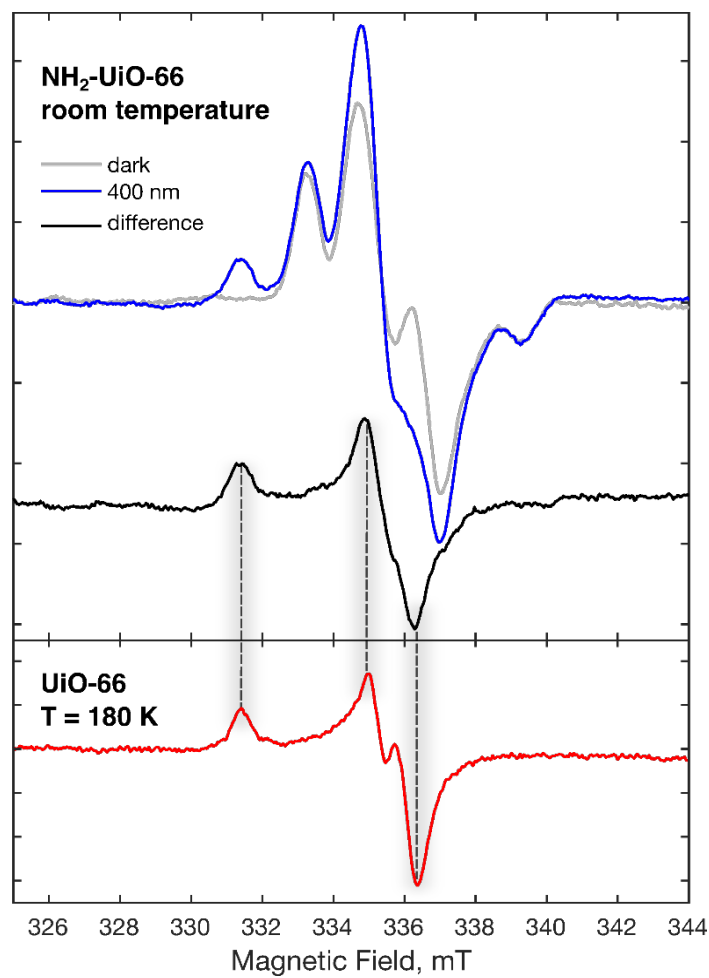

**Figure S6**. Comparison between the photoinduced EPR signal in NH$_2$-UiO-66 obtained at room temperature upon illumination with 400 nm (black line in *top subplot*) and the photoinduced EPR signal in pristine UiO-66 at 180 K (red line, *bottom subplot*) interpreted as a superoxide radical adsorbed at the SBU (for details, see Figure 3 and discussion in the main text). Since NH$_2$-UiO-66 is known to contain native spin centers, which are observed without illumination and are not photosensitive, the photoinduced signal was obtained as a difference between the illuminated (blue line) and dark (grey line) EPR spectra. The powder sample of NH$_2$-UiO-66 used in these measurements was synthesized according to the protocol described in Ref. [15] of the main text.

**Table S1.** Cartesian coordinates (in Å) of the cluster geometry used in the DFT calculations. Specific defect models can be introduced by performing constrained geometry optimization on this cluster, as described in the Methods section.

| Atom | X | Y | Z |
|---|---|---|---|
| C | 3.17958000 | 10.3502000 | 7.17062000 |
| C | 13.5297800 | 10.3502000 | 17.5208200 |
| C | 7.17062000 | 13.5297800 | 20.7004000 |
| C | 7.17062000 | 7.17062000 | 20.7004000 |
| C | 3.17958000 | 7.17062000 | 10.3502000 |
| C | 13.5297800 | 7.17062000 | 20.7004000 |
| C | 3.17958000 | 13.5297800 | 10.3502000 |
| C | 13.5297800 | 13.5297800 | 20.7004000 |
| C | 0.00000000 | 7.17062000 | 13.5297800 |
| C | 0.00000000 | 7.17062000 | 7.17062000 |
| C | 10.3502000 | 7.17062000 | 17.5208200 |

| C | 0.00000000 | 13.5297800 | 7.17062000 |
|---|---|---|---|
| C | 10.3502000 | 13.5297800 | 17.5208200 |
| C | 0.00000000 | 13.5297800 | 13.5297800 |
| C | 13.5297800 | 10.3502000 | 23.8799800 |
| C | 7.17062000 | 10.3502000 | 23.8799800 |
| C | -3.1795800 | 13.5297800 | 10.3502000 |
| C | 10.3502000 | 13.5297800 | 23.8799800 |
| C | -3.1795800 | 7.17062000 | 10.3502000 |
| C | 10.3502000 | 7.17062000 | 23.8799800 |
| C | -3.1795800 | 10.3502000 | 13.5297800 |
| C | -3.1795800 | 10.3502000 | 7.17062000 |
| Zr | 2.50665336 | 10.3439835 | 10.3332222 |
| Zr | 12.8753089 | 10.3506614 | 20.7061149 |
| Zr | 7.84351299 | 10.3439086 | 20.7173679 |
| Zr | -0.001611 | 12.8718958 | 10.3487212 |
| Zr | 10.3516646 | 12.8718561 | 20.7021087 |
| Zr | -0.0055824 | 7.8241483 | 10.3445041 |
| Zr | 10.3559403 | 7.82419678 | 20.705936 |
| Zr | -0.0170396 | 10.3439065 | 12.8568557 |
| Zr | -0.0056807 | 10.3506572 | 7.82508266 |
| Zr | 10.3671648 | 10.3441017 | 18.1936676 |
| Zr | -2.5250841 | 10.3506522 | 10.344401 |
| Zr | 10.3560256 | 10.350666 | 23.2254574 |
| C | 5.54543167 | 10.0346508 | 14.1718722 |
| C | 4.80480722 | 10.0420858 | 16.8782102 |
| C | 3.82136924 | 10.0421847 | 15.8955184 |
| C | 6.52859442 | 10.0340225 | 15.154714 |
| C | 7.21044179 | 10.137251 | 17.5604118 |
| C | 3.13968678 | 10.1376724 | 13.4901593 |
| C | 4.1919346 | 10.0528242 | 14.5419579 |
| C | 6.15846772 | 10.0526336 | 16.5081876 |
| O | 3.52355551 | 10.1879471 | 12.2839148 |
| O | 6.82634941 | 10.1903836 | 18.7662858 |
| O | 3.57852597 | 10.3948192 | 8.36507441 |
| O | 13.9327674 | 10.3563055 | 18.71519 |
| O | 1.98390455 | 13.9280793 | 10.3201121 |
| O | 12.3350455 | 13.9308073 | 20.6892186 |
| O | 1.98263952 | 6.7753708 | 10.3860421 |
| O | 12.3352812 | 6.76797417 | 20.7137304 |
| O | 8.36756191 | 6.77492466 | 20.667403 |
| O | 8.36602426 | 13.9288206 | 20.7277399 |
| O | 0.01075035 | 12.3334962 | 13.9305888 |
| O | 0.02179775 | 12.3349848 | 6.76879835 |
| O | 10.3375623 | 12.3336577 | 17.1196751 |
| O | -0.0290569 | 8.36450841 | 6.76884039 |
| O | 10.3955195 | 8.36157231 | 17.115208 |
| O | -0.0426979 | 8.36169782 | 13.9350928 |
| O | 0.0114809 | 13.9308299 | 8.3653439 |
| O | 10.3805922 | 13.9280203 | 18.7164492 |
| O | -0.0136378 | 6.76803239 | 8.36514107 |
| O | 10.3139433 | 6.77544928 | 18.717767 |
| O | 0.03336621 | 6.77503775 | 12.3328376 |
| O | -0.0282943 | 13.9284685 | 12.334268 |
| O | 3.58070327 | 12.3337024 | 10.3626428 |
| O | 13.9316304 | 12.3349934 | 20.6785098 |
| O | 6.77011216 | 12.3333831 | 20.6914792 |
| O | 6.76539359 | 8.36175118 | 20.7414338 |
| O | 3.58514673 | 8.36156237 | 10.3048584 |
| O | 13.931555 | 8.3645263 | 20.7294259 |
| O | 1.98523364 | 10.3562549 | 6.76757317 |
| O | 12.3353462 | 10.3945755 | 17.1218382 |
| O | 1.93389298 | 10.1894114 | 13.8741785 |
| O | 8.41638962 | 10.1880116 | 17.1766963 |
| O | 1.39295132 | 8.95014648 | 8.93717243 |
| O | 11.7631836 | 8.95017446 | 19.3074876 |
| O | 8.94925131 | 11.7374652 | 19.2996787 |
| O | 1.01368627 | 9.30990133 | 11.3639291 |
| O | 9.33659572 | 9.31012192 | 19.6868138 |
| O | 1.03225423 | 11.3951619 | 9.30301803 |
| O | 11.396197 | 9.31152832 | 21.7463674 |
| O | 11.7546429 | 11.7517653 | 22.1048204 |
| O | 13.9321029 | 10.3348165 | 22.6858056 |
| O | 6.7710998 | 10.3907372 | 22.6855728 |
| O | 9.31795793 | 11.3951444 | 21.7477648 |
| O | -1.0472917 | 11.3951045 | 11.3825112 |
| O | -1.984794 | 13.9306461 | 10.3634537 |
| O | 10.3359937 | 13.9304714 | 22.6851175 |
| O | -1.9850293 | 6.7683083 | 10.3341852 |
| O | -1.045872 | 9.31157424 | 9.3043339 |
| O | -1.412958 | 8.95005388 | 11.7432274 |
| O | 10.3666286 | 6.76830854 | 22.6854458 |
| O | -1.985149 | 10.392649 | 13.9289921 |
| O | -1.9853921 | 10.3345032 | 6.76834415 |
| O | 12.3355801 | 10.3323834 | 24.2821448 |
| O | 8.36510852 | 10.3591899 | 24.2825067 |
| O | -3.5816377 | 12.3350533 | 10.3704579 |
| O | 10.3310898 | 12.3351257 | 24.2822219 |
| O | -3.581633 | 8.36449068 | 10.3229857 |
| O | 10.3765203 | 8.36451798 | 24.2820584 |
| O | -3.5823987 | 10.3579723 | 12.3353747 |
| O | -3.5817662 | 10.3329682 | 8.36482124 |
| H | 14.3153617 | 6.38506957 | 20.6664198 |
| H | 14.3153985 | 10.3763024 | 24.6657084 |
| H | 14.3153338 | 14.3155761 | 20.7257166 |
| H | 6.38546345 | 14.3160643 | 20.684377 |
| H | 0.01527224 | 14.3163272 | 14.3146853 |
| H | 0.00806744 | 6.38030013 | 14.3113935 |
| H | 3.96426914 | 14.3165429 | 10.3655442 |
| H | 3.96096347 | 6.38006978 | 10.3585451 |
| H | -3.9649979 | 14.3156998 | 10.3244736 |
| H | -0.0254832 | 14.3155557 | 6.38505308 |
| H | 10.3348771 | 14.3165104 | 16.7361006 |
| H | -3.9648986 | 6.38484271 | 10.3849648 |
| H | -3.9652924 | 10.3761016 | 6.38498109 |
| H | 6.38897326 | 6.38032214 | 20.6934251 |
| H | 10.3158231 | 6.38484769 | 24.6653213 |
| H | 10.3422378 | 6.37998242 | 16.7395262 |
| H | 0.03428207 | 6.38502699 | 6.38509691 |
| H | 3.96652625 | 10.3105745 | 6.38679522 |
| H | -3.9635855 | 10.3110245 | 14.3165773 |
| H | 6.38405585 | 10.3116607 | 24.6642428 |
| H | 14.3136319 | 10.3107512 | 16.7338894 |
| H | 10.3757241 | 14.3158711 | 24.6652318 |
| H | 5.80697036 | 10.0364756 | 13.1127622 |
| H | 2.76236597 | 10.0509169 | 16.157121 |
| H | 7.58774837 | 10.0379851 | 14.8933566 |
| H | 8.38339285 | 12.2758327 | 18.7334119 |
| H | 12.3197546 | 8.3965643 | 18.7474951 |
| H | 12.3102573 | 12.3107843 | 22.6603882 |
| O | 8.95715244 | 8.94996324 | 22.1132542 |
| H | 8.39753854 | 8.39628118 | 22.6701269 |
| H | -1.9695295 | 8.39627177 | 12.3030438 |
| H | 1.95299412 | 8.39642357 | 8.38076706 |
| O | -1.4044096 | 11.7517588 | 8.94584595 |
| H | -1.9598983 | 12.3110001 | 8.39037419 |
| O | 1.40087975 | 11.7373019 | 11.751163 |
| H | 1.96684265 | 12.2756221 | 12.3173633 |
| O | 11.3973323 | 11.3951989 | 19.6681164 |
| H | 4.54331581 | 10.05167 | 17.9373332 |

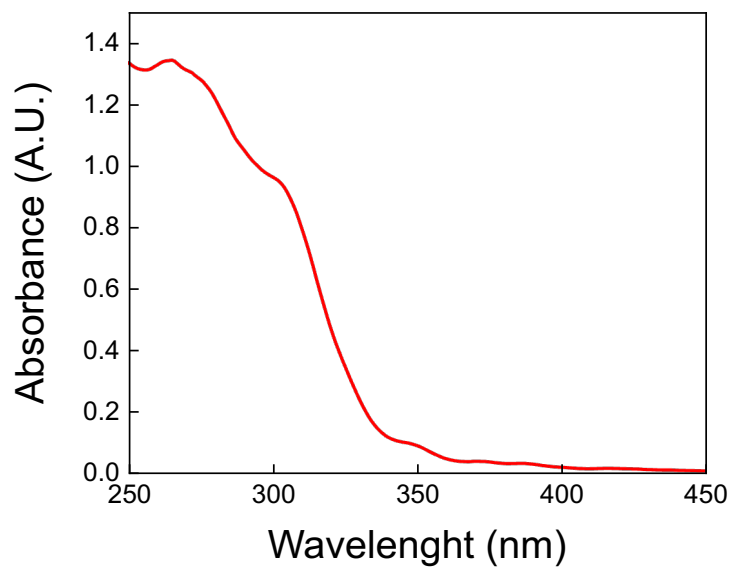

**Figure S7**. UV-visible DRS absorption spectra of UiO-66 samples.

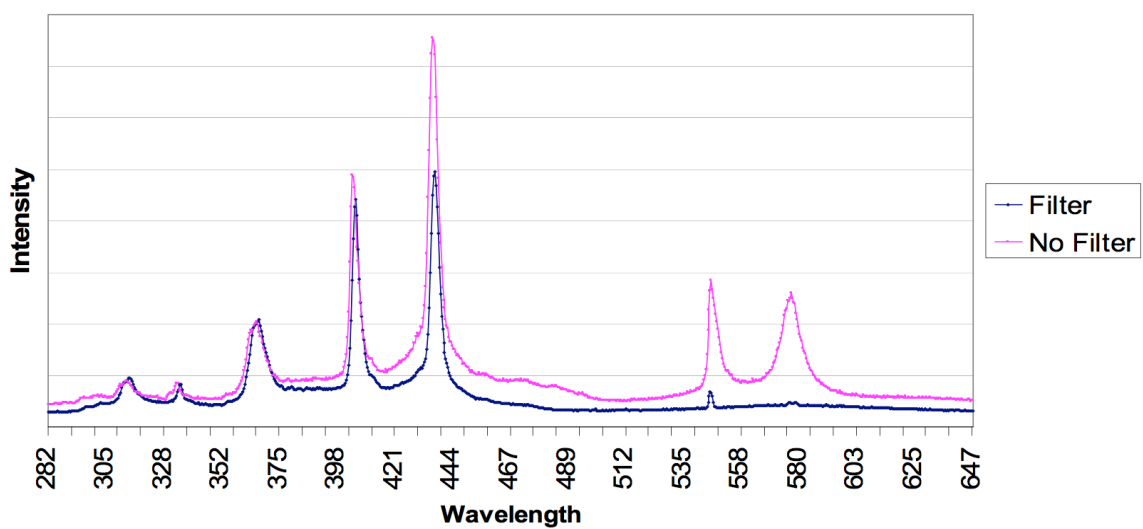

**Figure S8**. Spectrum of the used broadband Dymax BlueWave 50 UV lamp.

**REFERENCES**

1S. M. J. Katz, Z. J. Brown, Y. J. Colón, P. W. Siu, K. A. Scheidt, R. Q. Snurr, J. T. Hupp, K. O. Farha, *Chem. Comm.* 2013, **49**, 9449.

2S. S. Stoll, A. Schweiger, *J. Magn. Reson.* 2006, **178**(1), 42.

3S. S. Richert, A. E. Tait, C. R. Timmel, *J. Magn. Reson.* 2017, **280**, 103.

4S. V. Chechik, E. Carter, D. Murphy, Electron paramagnetic resonance, 2016, Oxford University Press.

5S. F. Neese, F. Wennmohs, U. Becker, C. Riplinger, *J. Chem. Phys.* 2020, **152**(22).

6S. V. N. Staroverov, G. E. Scuseria, J. Tao, J. P. Perdew, *J. Chem. Phys.* 2003, **119**(23), 12129.

7S. F. Weigend, *Phys. Chem. Chem. Phys.* 2006, **8**(9), 1057.

8S. F. Neese, F. Wennmohs, A. Hansen, U. Becker, *Chem. Phys.* 2009, **356**(1-3), 98.

9S. T. Petrenko, S. Kossmann, F. Neese, *J. Chem. Phys.* 2011, **134**(5), 054116.

10S. T. Biktagirov, W. G. Schmidt, U. Gerstmann, *Phys. Rev. Res.* 2020, **2**(2), 022024.

11S. S. Sinnecker, F. Neese, *J. Phys. Chem. A* 2006, **110**(44), 12267.

12S. J. P. Perdew, K. Burke, M. Ernzerhof, *Phys. Rev. Lett.* 1996, **77**(18), 3865.

Table of Contents Entry for

**Photoinduced Spin Centers in Photocatalytic Metal-Organic Framework UiO-66**

The paper explores the fundamental photophysical mechanisms in the UiO-66 metal-organic framework (MOF), an archetypical photocatalyst. Using Electron Paramagnetic Resonance (EPR) and time-resolved photoluminescence spectroscopy, the study identifies photoinduced spin centers, including a superoxide radical, an electron-hole pair, and a triplet exciton. The findings provide key insights into charge transfer processes underlying photocatalytic activity in MOFs.

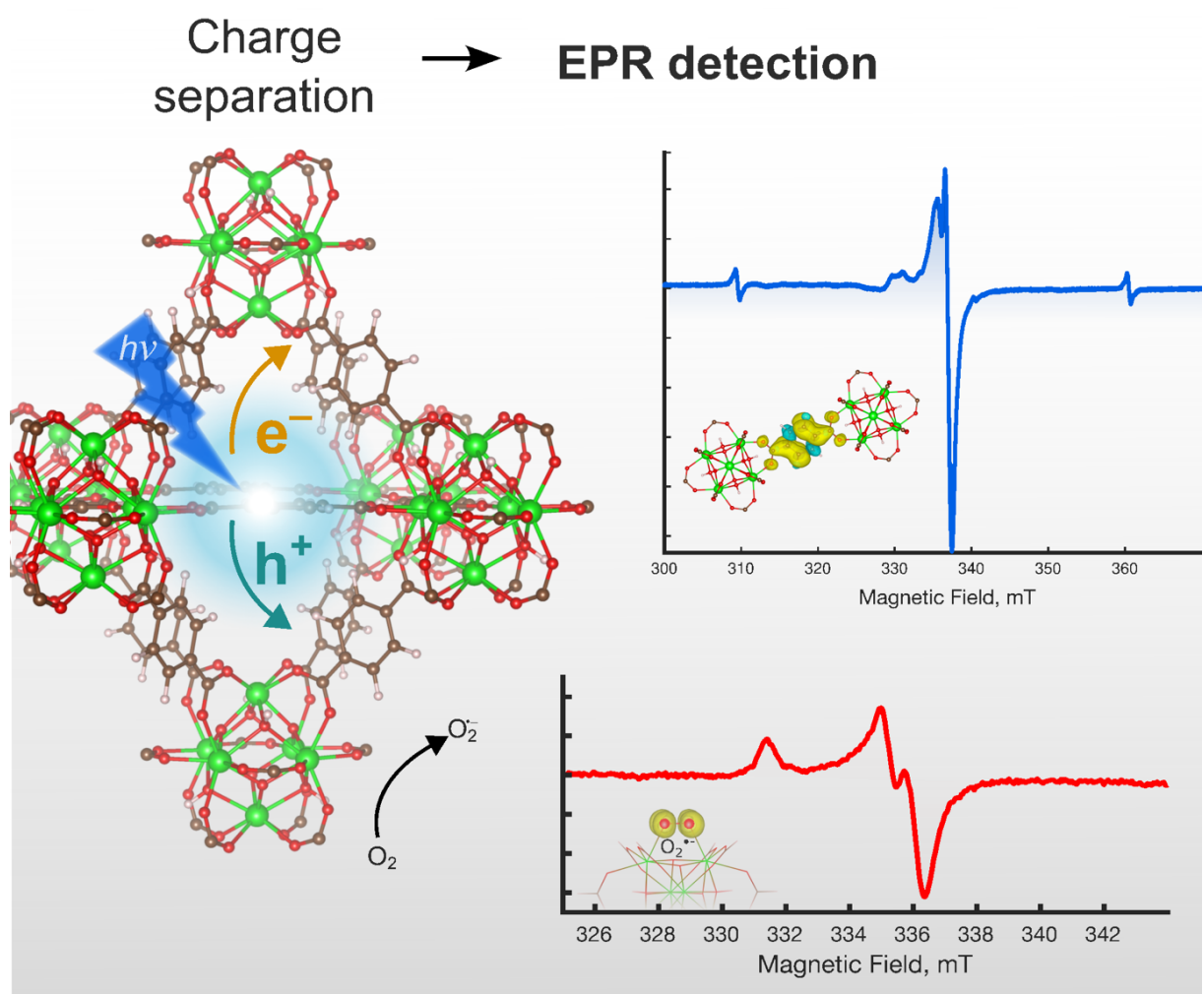